\documentclass[prb,aps,twocolumn,showpacs]{revtex4-1}

\usepackage{amsmath}
\usepackage{amssymb}
\usepackage{graphicx}
\usepackage[colorlinks=true,allcolors=blue]{hyperref}

\newcommand\pdag{{\phantom \dag}}
\newcommand\sh{\mathop{\rm sh}}

\begin{document}

\title{Second-order coherence of microwave photons emitted by a quantum point
contact}

\author{Fabian Hassler}
\author{Daniel Otten}
\affiliation{JARA-Institute for Quantum Information, RWTH Aachen University,
D-52056 Aachen, Germany}

\date{September 2015}

\begin{abstract} 
  Shot-noise of electrons that are transmitted with probability  $T$ through a
   quantum point contact (biased at a voltage $V_0$) leads to a fluctuating
   current that in turn emits radiation in the microwave regime. By
   calculating the Fano factor $F$ for the case where only a single channel
   contributes to the transport, it has been shown that  the radiation
   produced at finite frequency $\omega_0$ close to $e V_0/\hbar$ and at low
   temperatures is nonclassical with sub-Poissonian statistics ($F<1$). The
   origin of this effect is the fermionic nature  of the electrons producing
   the radiation, which reduces the probability of simultaneous emission of
   two or more photons.  However, the Fano factor, being a time-averaged
   quantity, offers only limited information about the system.  Here, we
   calculate the second-order coherence $g^{(2)}(\tau)$ for this source of
   radiation.   We show that due to the interference of two contributions, two
   photon processes (leading to bunching) are completely absent  at zero
   temperature for $T=50\,\%$.  At low temperatures, we find a competition of
   the contribution due to Gaussian current-current fluctuations (leading to
   bunching) with the one due to non-Gaussian fluctuations (leading to
   antibunching). At slightly elevated temperatures, the non-Gaussian
   contribution becomes suppressed whereas the Gaussian contributions remain
   largely independent of temperature. We show that the competition of the two
   contributions leads to a nonmonotonic behavior of the second-order
   coherence as a function of time.  As a result, $g^{(2)}(\tau)$ obtains a
   minimal value for times $\tau^* \simeq \omega_0^{-1}$. Close to this time,
   the second-order coherence remains below 1 at temperatures where the Fano
   factor is already above 1.   We identify realistic experimental
  parameters that can be used to test the sub-Poissonian nature of the
  radiation.
\end{abstract}

\maketitle

\section{Introduction}

In quantum optics, the degree of coherence plays a crucial role in
characterizing different sources of radiation. In particular, the second-order
coherence $g^{(2)}(\tau)$ is of central importance because it relates to the
statistics of the radiation.\cite{purcell:56} It can be shown that radiation
sources whose fluctuations are independent of the optical phase such as
lasers, thermal or chaotic light, all lead to $g^{(2)}(\tau) \geq 1$.
Microscopically, this result can be interpreted as an effect due to the
bunching of the photons. On the other hand, a radiation field with
$g^{(2)}(\tau)< 1$ indicates that the radiation cannot be described in
classical terms by a statistical superposition of coherent fields with
different intensities. It is of fundamental interest to find and characterize
radiation sources  that produce nonclassical light.  An important idea in this
respect is to use the fermionic nature of electrons emitting the photons to
imprint their antibunched statistics onto the radiation field. A prime example
are single photon sources where a single electron in an excited state of an
atom is used to emit a single photon.\cite{lounis:05}  An obvious question in
this direction is whether there is a quantum limit to the classical light bulb
where a resistive wire that is biased with a \emph{DC}-voltage is employed to
produce photons.  This question has been affirmatively
answered:\cite{beenakker:04} the idea is to use a quantum point contact biased
by a voltage $V_0$. Single electrons are then transmitted stochastically (with
probability $T$) through the barrier leading to current fluctuations. The
photons produced by these current fluctuations are nonclassical provided the
system is kept at low temperatures $\vartheta \ll eV_0/k_B$, with $e>0$ the
elementary charge and $k_B$ the Boltzmann constant, such that the electron
reservoirs are degenerate and bunching of the photons is suppressed.  The
suppression of bunching is achieved by two measures: firstly, only a single
channel is allowed to contribute to the transport which requires the breaking
of the spin-degeneracy in a magnetic field.  Secondly, by engineering of the
environment, the photons should be emitted preferentially at frequencies
$\omega \geq eV_0 /2\hbar$ such that each electron that carries an energy less
than $eV_0$ emits at most a single photon.

\begin{figure}[tb]
  \centering
  \includegraphics{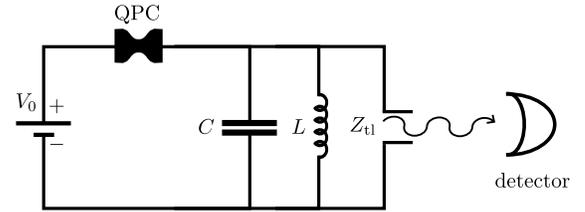}
  \caption{%
    Source (left part) and analyzer (right part) of the nonclassical
    radiation: A quantum point contact (QPC) is biased at a constant voltage
    $V_0$ and produces current-current fluctuations due to shot and thermal
    noise. These fluctuations are filtered by an \emph{LC}-resonator and transmitted
     to the detector as electromagnetic-radiation via a transmission line with
    characteristic impedance $Z_\text{tl}$.
  }\label{fig:setup}
\end{figure}

The long-time properties of the radiation emitted by a quantum point contact,
such as the Fano factor and even arbitrary cumulants of the photon counting
statistics, have been obtained both at zero temperature
\cite{beenakker:01,beenakker:04,lebedev:10} and  at finite temperatures
\cite{fulga:10}. Based on ideas of Ref.~\onlinecite{lesovik:97}, a detector
formed by two resonators at different frequencies has been analyzed in
details.\cite{lebedev:10} Experimental progress has been impressive, resulting
in the measurement of high-frequency shot-noise \cite{zakka-bajjani:07} and
the measurement of the two photon interference of photons emitted by a tunnel
junction \cite{zakka-bajjani:10}. One of the last remaining challenges, the
impedance matching of the wave-guides transporting the radiation from the
quantum point contact to the amplifier that acts as a detector, has been
recently resolved in employing superconducting
circuits.\cite{hofheinz:11,altimiras:13} Here, we want to study the
second-order coherence $g^{(2)}(\tau)$, which is a time-resolved quantity and
thus also carries information about the correlation time of the source. In
particular, we show that the behavior of $g^{(2)}(\tau)$ as a function of the
autocorrelation time $\tau$ is nonmonotonous with a minimum around $\tau^*
\approx 5\omega_0^{-1}$. We highlight that a value $g^{(2)}(\tau^*) <1$ can be
obtained at temperatures where the radiation is already super-Poissonian.

The outline of the paper is as follows: In Sec.~\ref{sec:setup}, we present
the experimental setup that is  analyzed in the remainder of the paper.
Sections~\ref{sec:nav} and \ref{sec:n2} provide the general results for the
average photon rate as well as the photon-photon correlator that is directly
linked to the second-order coherence. In Sec.~\ref{sec:zero}, we provide
analytical results valid at zero temperature.  We proceed by determining the
critical temperatures  below which the observation of nonclassical features in
the radiation is possible in time-averaged quantities (Sec.~\ref{sec:tc}) as
well as the second-order coherence (Sec.~\ref{sec:tcs}). We conclude in
Sec.~\ref{sec:experimental} by identifying a set of realistic experimental
parameters and provide results for the Fano factor and the second-order
coherence for this case.

\section{Setup \& Model}\label{sec:setup}

The setup we have have in mind is shown in Fig.~\ref{fig:setup}. The quantum
point contact is assumed to be voltage-biased at a voltage $V_0$ and thus
produces current noise $e^2 S(\omega) = \int\!dt\,\langle \langle  I(0) I(t)
\rangle\rangle e^{i\omega t}$ at frequency $\omega$ where $I(t)$ denotes the
current through the quantum point contact and $\langle\langle \cdot
\rangle\rangle$ indicates a cumulant.  In the following, we assume that only a
single channel with an energy-independent transmission probability $0\leq T
\leq 1$ contributes to the transport as the nonclassical signatures of the
radiation are absent two or more channels.\cite{Note1}

The source of current noise is embedded in an electromagnetic environment
formed by an \emph{LC}-resonator and a wide-band transmission line in
parallel. The \emph{LC}-resonator is characterized by the resonance frequency
$\omega_0 = (L C)^{-1/2} $ and the characteristic impedance $Z_0 =
(L/C)^{1/2}$, which is the ratio of the voltage to the current in the
\emph{LC}-resonator at the resonance frequency. As we want to treat the
quantum point contact with impedance $2 \pi \hbar /e^2 T$ to be biased by a
constant voltage, we have to assume that the impedance of the environment is
small compared to the impedance of the quantum point contact, that is $G_Q Z_0
\ll 1$ with $G_Q =e^2/2\pi \hbar$.\cite{ingold:92} The transmission line
introduces both damping as well as a slight shift of the resonant frequency.
The latter can be incorporated in a redefinition of $\omega_0$. The only
relevant contribution of the impedance of the transmission line is its real
part $Z_\text{tl}(\omega \approx \omega_0) = Z_0 \omega_0/\gamma$ that we will
parameterize by the rate $\gamma$ whose meaning will be made clear below.  

As the inductance, capacitance, and transmission line form a parallel circuit,
the total impedance $Z_\omega = V_\omega/I_\omega$, relating the current $I(t)
= \int(d\omega/2\pi) I_\omega e^{-i\omega t}$ through the quantum point
contact to the voltage $V(t)$ across the \emph{LC}-resonator, is given by
\begin{equation}\label{eq:z}
  Z_\omega = \frac{ Z_0 \omega_0 \omega}{ i (\omega_0^2 -\omega^2)
  +\gamma \omega}.
\end{equation}
As described in the introduction, the density of states for photons in the
environment should be large for photons with frequency larger than
$eV_0/2\hbar$. In order to achieve that, we have to require that $\gamma
\lesssim eV_0/2\hbar$ and $\omega_0 \gtrsim eV_0/\hbar$. We have found that
the optimal value of $\omega_0$ is close to $e V_0/\hbar$, especially at
temperatures close to the critical temperature. In order to keep the number of
parameters manageable, we thus present only results for the resonant condition
$\hbar\omega_0 =eV_0$ in the following (except for Fig.~\ref{fig:volt_tau}).

For positive frequencies (emission), the impedance is well-approximated by the
form
\begin{equation}\label{eq:zt}
  \tilde Z_\omega = \frac{Z_0 \omega_0}{ 2 i (\omega_0 -\omega) + \gamma}
\end{equation}
that corresponds to the response of an oscillator at frequency $\omega_0$ with
quality factor $Q= \omega_0/\gamma$. In particular, the rate $\gamma$
describes the loss of photons from the cavity into the transmission
line.\cite{padurariu:12} Note that the total impedance on resonance
$Z_{\omega_0} = Q Z_0$ is a factor $Q$ larger than the characteristic
impedance of the resonator. 

In a next step, we want to find an expression for the rate of photons $n(t)$
that are emitted at time $t$ into the transmission line as those photons will
be subsequently detected in the amplifier  positioned at the other end. The
photon number $dn (t) = P_\omega(t) d\omega/2\pi\hbar \omega$ in the frequency
interval $[\omega, \omega+d\omega]$ is given by the power loss $P_\omega(t)=
\mathop{\rm Re} (Z^{-1}_\omega)  \int\! d\delta\, V(t-\delta/2) V(t+\delta/2)
e^{i\omega \delta}$ at this frequency measured in units of the photon energy
$\hbar \omega$.  Expressing the voltage $V_\omega =Z_\omega I_\omega$ via the
current, we arrive at the important result
\begin{equation}\label{eq:n_t}
n(t) =\iint\! \frac{d\omega \,d\nu}{2\pi e^2} 
  \alpha_\omega
   Z^*_{\omega+\nu/2} Z_{\omega-\nu/2}
   I^{-}_{\omega+\nu/2} I^+_{\omega-\nu/2} e^{i\nu t}
\end{equation}
with $\alpha_\omega= G_Q/Z_0 Q \omega$, $I^+_\omega =I_\omega \Theta(\omega)$
the current  projected on the positive frequency contributions, and
$I^-_\omega = (I^+_\omega)^\dag$;\cite{Note2} here and below  $\Theta(x)$
denotes the unit-step function.  Positive frequency in this context
corresponds to photon emission processes where energy from the electronic
system  is converted into photons.  Note that in the limit of large quality
factors $Q\gg1$ when $Z_\omega$ is well-approximated by $\tilde Z_\omega$, the
photons are solely emitted at the frequency $\omega_0$ such that
$\alpha_\omega$ can be approximated by $\tilde \alpha_\omega =  G_Q/Z_0 Q
\omega_0$, which is independent of $\omega$.

Below, the current $I(t)$ will be promoted to an operator and the usual
normal-ordering prescription of photon counting denoted by colons will be
assumed which implies that the `$+$' operators are positioned to the right of
the `$-$' operators, with the `$+$' operators being time-ordered, and the
`$-$' operators anti-time-ordered among themselves.\cite{Note3} Using this
notation, the main task of this paper is the evaluation of the second-order
coherence defined as
\begin{equation}\label{eq:g2}
  g^{(2)}(\tau)   = \frac{\langle \mathopen{:} 
  n(\tau) n(0) \mathclose{:} \rangle}{\langle n \rangle ^2}.
\end{equation}

The current through the quantum point contact is the source of the radiation
that is emitted in the transmission line. It can be evaluated using the
conventional Landauer-B\"uttiker approach of transport. We model the quantum
point contact by two electronic reservoirs, one  to the left and one to the
right of the constriction. The current operator at frequency $\omega$ has the
explicit form $I_\omega= I_{\text{out},\omega} -I_{\text{in},\omega} $ with
\begin{align}\label{eq:current}
  I_{\text{in},\omega} = e\!\int\!d\epsilon\, c^\dag_{R,\epsilon}
  c^\pdag_{R,\epsilon + \omega} , \quad
 I_{\text{out},\omega} =  e\!\int\!d\epsilon\,d^\dag_{\epsilon}
  d^\pdag_{\epsilon + \omega};
\end{align}
here, $d_\epsilon=  T^{1/2} c_{L,\epsilon} - R^{1/2} c_{R,\epsilon}$ with the
reflection probability is given by $R=1-T$ due to unitarity.  The electronic
states in the reservoirs are described by fermionic operators $c_{x,\epsilon}$
that fulfill the canonical anticommutation relations $\{c_{x,\epsilon},
c_{x',\epsilon'} \} = 0$ and  $\{c^\pdag_{x,\epsilon}, c^\dag_{x',\epsilon'}
\} = \delta_{xx'} \delta(\epsilon -\epsilon')$  with $x,x'\in\{L,R\}$ . The
states are assumed to be in (local) equilibrium and thus occupied according to
the Fermi-Dirac distribution $\langle c^\dag_{x,\epsilon}
c^\pdag_{x',\epsilon'} \rangle = \delta_{xx'} f_{x}(\epsilon)
\delta(\epsilon-\epsilon')$ with
\begin{equation}
  f_{x}(\epsilon) = \bigl(\exp[(\hbar
\epsilon -\mu_{x})/k_B \vartheta] +1 \bigr)^{-1}.
\end{equation}
In the following, we will measure the electronic energies with respect to the
chemical potential of the right reservoir and thus set $\mu_R= 0$ and $\mu_L
=e V_0$ (due to the voltage bias).

\section{Average Photon Rate}\label{sec:nav}

\begin{figure}[tb]
  \centering
  \includegraphics[width=0.8\linewidth]{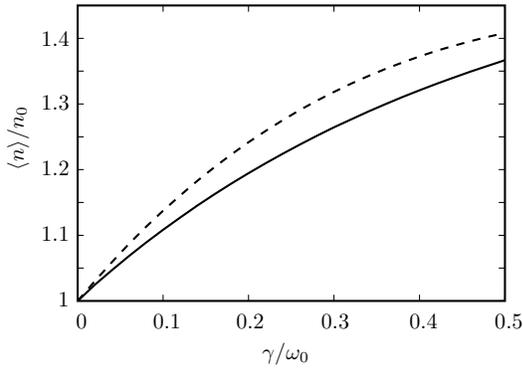}
  \caption{Comparison of the photon rate of Eq.~\eqref{eq:nav} (solid line)
  with the approximate expression Eq.~\eqref{eq:nav0} (dashed line) for
  $T=\tfrac12$ and $k_B \vartheta/e V_0 = 0.1$ as a function of
  $\gamma/\omega_0$ normalized to the value $n_0= R T G_Q Z_0 k_B
  \vartheta/4\hbar$ valid for $\gamma \to 0$.}\label{fig:nav}
\end{figure}

The photon rate can be calculated by averaging the expression \eqref{eq:n_t}
over the distribution of the electrons. Due to the fact that
the bias $V_0$ is constant, the state is stationary and as a result the
average photon rate $\langle n(t) \rangle$ is independent of $t$.  The
photon rate assumes the simple form
\begin{equation}\label{eq:nav}
  \langle n \rangle = \int_0^\infty \!d\omega \, \alpha_\omega
  |Z_\omega|^2 S(\omega).
\end{equation}
The current noise power $S(\omega)$ can be evaluated by employing
Wick's theorem yielding  the result $S (\omega) = S_\text{ex}(\omega)
+ S_\text{th}(\omega)$ with 
\begin{align}
  S_\text{ex}(\omega) &=RT\int\!\frac{d\epsilon}{2\pi}\,
  \Delta(\epsilon+\omega) \,\Delta(\epsilon)  \\
  &=  \frac{RT  \sh(\frac{\hbar\omega_0}{2k_B \vartheta})}{4\pi\sh (
     \frac{\hbar\omega}{2k_B \vartheta})} 
     \biggl( \frac{\omega_0 -\omega}{\sh\bigl[
       \frac{\hbar(\omega_0-\omega)}{2 k_B
     \vartheta} \bigr]}
    - \frac{\omega_0+\omega}{\sh\bigl[\frac{\hbar(\omega_0+\omega)}{2 k_B
    \vartheta}\bigr]}
  \Biggr) \nonumber\\
  \intertext{and}
  S_\text{th}(\omega)&=
  T \int\!\frac{d\epsilon}{\pi}  f_R(\epsilon+\omega) f_R(-\epsilon) 
  =\frac{T \omega}{\pi(e^{\hbar\omega/k_B \vartheta}- 1)}; \nonumber
\end{align}
here, we have introduced the abbreviation  $\Delta (\omega) = f_{L}(\omega) -
f_R (\omega)$ for convenience. The term $S_\text{th}(\omega)$ in the
expression for $S(\omega)$ describes the thermal noise whereas
$S_\text{ex}(\omega)$  is the excess noise due to the application of a finite
bias $V_0$. At zero temperature, only the excess noise remains at positive
frequencies. Note that at low temperatures and large quality factors, we may
replace $S$, $Z$, and $\alpha$ in \eqref{eq:nav} by $S_\text{ex}$, $\tilde Z$,
and $\tilde \alpha$ respectively. As a result, we obtain the approximate
expression
\begin{equation}\label{eq:nav0}
\langle n\rangle =\int_0^\infty \!d\omega\,
\frac{G_Q Z_0 \gamma S_\text{ex}(\omega)}{4 (\omega_0 -
\omega)^2 + \gamma^2}.
\end{equation}
Figure~\ref{fig:nav} shows a comparison of the photon rate,
Eq.~\eqref{eq:nav}, with the approximation \eqref{eq:nav0}.

The photon rate depends on the three energy scales $eV_0 $, $\hbar\gamma$, and
$k_B \vartheta$. We can evaluate $\langle n \rangle$ in the different relevant
limits and obtain the following approximations
\begin{equation}\label{eq:napp}
  \langle n \rangle 
  \approx \frac{R T G_Q Z_0}{8\pi}  \begin{cases} 
    2 \pi  k_B \vartheta/\hbar, & \hbar\gamma\ll k_B \vartheta \ll eV_0, \\
    \gamma\ln (2 Q)          ,  & k_B \vartheta \ll \hbar\gamma \ll eV_0.
  \end{cases}
\end{equation} 
Since  $\gamma$ denotes the rate at which photons are transferred from the
\emph{LC}-resonator to the transmission line, the number of photons in the
cavity $n_\text{cav}$ is given by $\langle n \rangle /\gamma \simeq RT G_Q Z_0
\max(1, 2\pi k_B \vartheta/\hbar \gamma)/8\pi$. As we are mainly interested in
the low-temperature limit in the following, we use the abbreviation
\begin{equation}\label{eq:n_cav}
  n_\text{cav} = \frac{R T G_Q Z_0}{8\pi}
\end{equation}
consistently.

\section{Photon-photon correlation}\label{sec:n2}

In order to obtain the second-order coherence, the task is to obtain the
photon-photon correlation function $n^{(2)}(\tau)=\langle \langle \mathopen{:}
n(\tau) n(0) \mathclose{:} \rangle \rangle =\langle \mathopen{:} n(\tau) n(0)
\mathclose{:}\rangle - \langle n\rangle^2$. Given this correlation function,
the second-order coherence follows via
\begin{equation}\label{eq:g2_n2}
  g^{(2)}(\tau) = \frac{n^{(2)}(\tau)}{\langle n \rangle^2} +1.
\end{equation}
In particular, a negative correlation function $n^{(2)}(\tau)$ at some time
$\tau$ is equivalent to having $g^{(2)}(\tau)<1$ and thus indicates
nonclassical radiation. The prior work
\cite{beenakker:01,beenakker:04,lebedev:10} has been concentrated on obtaining
zero frequency (long-time) results. In particular, the Fano factor $F=
\int\!d\tau\,\langle\langle \mathopen{:} n(\tau) n(0) \mathclose{:}
\rangle\rangle / \langle n \rangle$ describing whether a source is more
correlated ($F>1)$ or anticorrelated ($F<1$) than a Poissonian source with
$F=1$ has been of central interest. The connection with $g^{(2)}$ is provided
by\cite{emary:12}
\begin{equation}\label{eq:g2_F}
 F - 1 = \langle n \rangle \int\!d\tau\, [g^{(2)}(\tau) -1]
 = \frac{N^{(2)}}{\langle n \rangle} .
\end{equation}
where we have introduced the total number of correlated photons $N^{(2)} =
\int\!d\tau \,n^{(2)}(\tau)$.  A Fano factor with $F<1$ is a clear indication
that $g^{(2)}(\tau)$ is smaller than one for some $\tau$. However, having the
full information $g^{(2)}(\tau)$ available, it is possible to have
$g^{(2)}(\tau) < 1$ for some time $\tau = \tau^*$ even when $F\gtrsim 1$.

\begin{figure}[tb]
  \centering
  \includegraphics{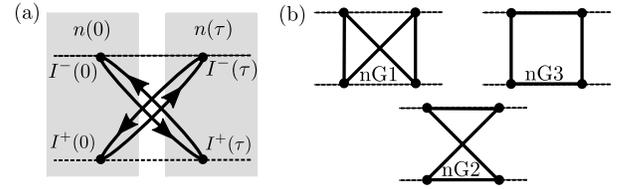}
  \caption{%
    Diagrams contributing to the photon-photon correlator $n^{(2)}$.  The
    dashed lines denote the Keldysh contour and the current operators $I^{+}$
    ($I^{-}$), denoted by black dots, are located on the lower (upper) branch.
    Each photon-number operator $n$  (gray boxes) consists of a current
    operator on each branch.   (a) The contribution due to  Gaussian
    current-current fluctuations:     Each current operator consists of two
    fermionic operators and the solid lines with the arrows indicate their
    contractions.  Although the diagram is reducible in terms of the current
    operators [as it decays into a product of two second-order correlators
    $S(\omega)$], it contributes the genuine term \eqref{eq:n2d}  to the
    irreducible correlator $n^{(2)}(\tau)$.  (b) The three diagrams (nG1, nG2,
    nG3) of the non-Gaussian contributions to the photon-correlation function
    that originate from irreducible forth-order current correlators.  The
    sketches are analogous to (a): The current-operators are shown as black
    dots, the Keldysh contour as dashed lines, and the contractions as solid
    lines. We have omitted the arrows on the solid lines as each diagram
    contributes twice with the contributions differing by the direction of the
    arrows that define the contractions.\cite{otten}
  }\label{fig:diag}
\end{figure}

Obtaining the photon-photon correlation function $n^{(2)}(\tau)$ is more
difficult than the average photon rate.  The problem is that it involves the
evaluation of a forth-order current correlators.  We proceed with the insight
of Ref.~\onlinecite{beenakker:01} that the normal-ordering of the current
operators is equivalent to the \emph{in-out}-ordering, \emph{i.e.}, ordering
$I^-_\text{in}$, $I^-_\text{out}$, $I^+_\text{out}$, $I^+_\text{in}$ from left
to right. The concrete ordering within a group of these operators does not
matter, since $[I_\text{in}(t), I_\text{in}(t') ]=0$ and similarly for
$I_\text{out}$.\cite{Note4} The evaluation of $n^{(2)}$ thus proceeds in the
following three steps: ($i$) We express $n(t)$ via the current operators using
the general expression \eqref{eq:n_t}. ($ii$) We implement the normal-ordering
for each term utilizing the prescription in terms of the
\emph{in-out}-ordering as explained above.  ($iii$) In each term, we introduce
the expression \eqref{eq:current} for the current operators and then perform
the average over the reservoir with the help of Wick's theorem for the
electronic operators $c_{L/R}$. The relevant diagrams for calculating
$n^{(2)}(\tau)$ are shown schematically in Fig.~\ref{fig:diag}.

The photon-photon correlation function can be decomposed into two parts
$n^{(2)}(\tau) = n^{(2)}_\text{G}(\tau) + n^{(2)}_\text{nG}(\tau) $ with a
Gaussian term, which dominates at large temperatures and is always positive,
given by
\begin{multline}\label{eq:n2d}
  n^{(2)}_\text{G}(\tau) =
  \iint_0^{\infty}\!d\omega_1 d\omega_2  \,
  \alpha^2_{(\omega_1  + \omega_2)/2}  |Z_{\omega_1}|^2 |Z_{\omega_2}|^2 \\ 
  \times S(\omega_1) S(\omega_2)
  \cos[(\omega_1 - \omega_2)
  \tau]
\end{multline}
and a non-Gaussian term
\begin{multline}\label{eq:n2e}
  n^{(2)}_\text{nG}(\tau)= 
   \iint \! 
  \frac{d\nu d\epsilon}{(2\pi)^2}
  \iint_{|\nu|/2}^\infty\! \!
  d \omega_1 d\omega_2 \, Z^{(2)} (\omega_1, \omega_2,\nu) \\
  \times( n_\text{nG1} + n_\text{nG2} + n_\text{nG3}) \cos(\nu \tau),
\end{multline}
which  depends through the combination
\[
  Z^{(2)}= \alpha_{\omega_1}
  \alpha_{\omega_2} \mathop{\rm Re} \bigl( Z_{\omega_1 -\nu/2} Z_{\omega_2 +\nu/2}
Z_{\omega_1+\nu/2}^* Z_{\omega_2 -\nu/2}^* \bigr)
\]
on the impedance $Z_\omega$ such that $n^{(2)}_\text{nG}$ different from
$n^{(2)}_\text{G}$ depends on the phase of $Z_\omega$. More importantly, the
non-Gaussian term does not have a well-defined sign. In fact it has been shown
that if the impedance is peaked in the frequency interval $[eV_0/2\hbar,
eV_0/\hbar]$ and at low temperatures, the non-Gaussian contribution is negative
and for proper choice of parameters even dominates the direct
contribution.\cite{beenakker:04}

The non-Gaussian contribution involves the three diagrams nG1, nG2, and
nG3.  The general expressions are quite involved though they follow
straightforwardly from the  recipe outlined above. At zero temperature, only a
few terms survive and the resulting contributions can be written in the
compact form
\begin{align}\label{eq:nej}
  n_\text{nG1} &\!=\! -2 R^2 T^2  \Delta (\epsilon\!-\!\tfrac12\nu) \,\Delta
  (\epsilon+\tfrac12\nu) \,\Delta (\epsilon+\omega_1) \,\Delta
  (\epsilon+\omega_2), \nonumber\\
  n_\text{nG2} &\!=\! -2 R^2 T^2 \Delta (\epsilon)  
  \Delta (\epsilon + \omega_1 +\tfrac12 \nu) \Delta (\epsilon + \omega_2
+ \tfrac12 \nu)  \nonumber\\
  &\qquad \times
 \Delta (\epsilon+\omega_1 +\omega_2), \nonumber\\
  n_\text{nG3} &\!=\!RT (1-2 RT)
   \Delta (\epsilon) 
   \Delta (\epsilon + \omega_2 \!-\!\tfrac12 \nu)   \Delta (\epsilon + \omega_2
+ \tfrac12 \nu)   \nonumber\\
   &\qquad \times
 \Delta (\epsilon+\omega_1 +\omega_2) ,
\end{align}
where $\Delta(\epsilon)$ selects electrons with energies $\epsilon$ within the
transport window. In the following, we use the approximate expressions for
$n_\text{nG$j$}$ also at finite temperatures. We have tested numerically that
for low temperatures $\vartheta \lesssim eV_0/k_B$, which we are interested
in, the results differ from the exact expression by not more than a few
percent, see for example the dashed lines in Figs.~\ref{fig:tc} and
\ref{fig:g2}.  The analytical expressions in Eq.~\eqref{eq:nej} are one of the
central results of this paper as they provide  an accurate analytical
description of the physics which we want to discuss in the following.

From the three contributions, $n_\text{nG1}$ and $n_\text{nG2}$ are negative
and thus lead to antibunching whereas $n_\text{nG3}$ is positive.  The three
contributions have different physical origin. The first contribution
$n_\text{nG1}$ originates from a correlated emission of two photons  at
frequency $\omega_1$ and $\omega_2$ due to the transfer of two electrons. For
$\nu=0$, this term has already been discussed in Ref.~\onlinecite{fulga:10}.
The other two terms are new and describe two photon processes where a single
electron which is transmitted through the quantum point contact emits two
photons. Note that these processes are suppressed for an environment such that
the impedance for $\omega < e V_0/2\hbar$ is vanishingly small. In our setup,
the smallness of the impedance in this regime is controlled by the quality
factor $Q$.

The evaluation of $N^{(2)}$, which is needed for the Fano factor, involves the
regime of long-measurement time. The integral over $\tau$ in \eqref{eq:g2_F}
then reduces the number of frequency integrals in the expressions of
$n^{(2)}(\tau)$ by one as $\int\!d\tau\,\cos(\nu \tau) =2 \pi \delta(\nu)$.
Since the main negative contribution $n_\text{nG1}$ to $n^{(2)}$ is largest for
$T=\tfrac12$, we will present only results for this case in the following.

\section{Zero temperature}\label{sec:zero}

We first present the results at vanishing temperatures. In this case, the
Fermi-Dirac distribution becomes a step function. The physics is concentrated
on energies within the voltage bias with $\Delta(\epsilon) = 1$ for $\epsilon
\in [0,eV_0]$ and zero otherwise.  In the direct contribution to the
photon-photon correlator, we  can replace the shot-noise $S(\omega)$ by the
zero temperature result  $ RT (\omega_0 -\omega) \Theta(\omega_0
-\omega)/2\pi$ and obtain
\begin{multline}\label{eq:n_d0}
  n^{(2)}_\text{G} (\tau) = \frac{R^2 T^2}{4\pi^2} \iint_{0}^{\omega_0}
 d\omega_1\,d\omega_2\,
  (\omega_0 - \omega_1)(\omega_0 -\omega_2) \\
  \times \alpha^2_{(\omega_1 + \omega_2)/2}
  |Z_{\omega_1}|^2  |Z_{\omega_2}|^2  \cos[(\omega_1 - \omega_2)\tau]
\end{multline}

Since $\Delta(\epsilon)$ is a step function, we can also simplify the
expression for $n_\text{nG$j$}$ with the results
\begin{align}\label{eq:f_s0}
  n_\text{nG1} &=  -2 R^2 T^2\Delta( \epsilon - \tfrac12 |\nu|) \Delta(\epsilon +
  \omega_1) \Delta(\epsilon+ \omega_2), \nonumber\\
  n_\text{nG2} &= -2 R^2 T^2 \Delta(\epsilon) \Delta(\epsilon +
  \omega_1 + \omega_2), \nonumber\\
  n_\text{nG3} &=  RT (1-2 RT)
   \Delta (\epsilon) 
 \Delta (\epsilon+\omega_1 +\omega_2) .
\end{align}
For $T=\tfrac12$, which is the optimal choice to observe the photon
antibunching, we have $n_\text{nG2} + n_\text{nG3}=0$ and two photon processes
are absent irrespective of the shape of $|Z_\omega|$. As, this feature will
approximately persist also to some small but finite temperatures, the
stringent requirements on the quality factor $Q$ of the cavity can be relaxed.
The $\epsilon$ integration can now be performed readily with the result (valid
for $T=R=\tfrac12$)
\begin{multline}\label{eq:n_e0}
  n_\text{nG}^{(2)}(\tau) = - \frac{2R^2 T^2}{\pi^2}
  \int_\mathcal{R}\!d\nu\,d\omega_1\,d\omega_2 \, (\omega_{0} -\tfrac12 \nu
  - \omega_2)   \\\times
  Z^{(2)}(\omega_1, \omega_2 ,\nu) \cos(\nu \tau)
\end{multline}
where the integration is constraint onto the region $\mathcal{R}$ with
$0<\nu/2 < \omega_1 < \omega_2 < \omega_0 - \nu/2$.

The question whether the radiation can be classified as nonclassical relies on
the competition between the positive (classical) contribution \eqref{eq:n_d0}
and its negative counterpart \eqref{eq:n_e0}. As explained before, for the
evaluation of $N^{(2)}$ the $\cos$-factor becomes a $\delta$-function and thus
reduces the number of integrals by one. Solving the remaining integrals yields
the results (valid for $Q \gg 1$)
\begin{align*}\label{eq:N_2}
  N_\text{G}^{(2)} &= \pi^2 n_\text{cav}^2 \gamma,&
  N_\text{nG}^{(2)} &=  -8\ln(2) \pi^2  n_\text{cav}^2 \gamma.
\end{align*}
Because $|N^{(2)}_\text{nG}|/N^{(2)}_\text{G} =8\ln 2 \approx 5.5 > 1$ the
Fano factor is below 1 and we obtain
\begin{equation}\label{eq:fano}
  F - 1 = \frac{(1- 8\ln 2) \pi^2 n_\text{cav}}{\ln (2Q)}, \qquad
  Q \gg 1.
\end{equation}
\begin{figure}[tb]
  \centering
  \includegraphics[width=.7\linewidth]{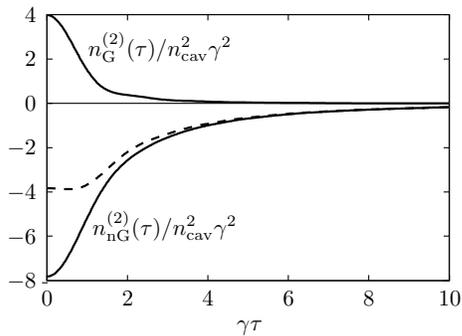}
  \caption{%
    Photon correlation functions $n^{(2)}_\text{G}(\tau)$ due to Gaussian
    current-current fluctuations and $n^{(2)}_\text{nG}(\tau)$ due to
    non-Gaussian fluctuations at zero temperature for $Q=5$. It can be seen
    that both correlators decay on a characteristic scale set by the inverse
    of the cavity decay rate. The positive contribution $n^{(2)}_\text{G}$ and
    the negative contribution $n^{(2)}_\text{nG}$ add up to the photon-photon
    correlator $n^{(2)}(\tau)$ (dashed line). As $n^{(2)}$ is negative, the
    second-order coherence $g^{(2)}(\tau)$ is smaller than 1 indicating that
    the photons are antibunched.
  }\label{fig:n2}
\end{figure}

Next, we turn to the discussion of the photon-photon correlator. For
$\tau=0$, we obtain the result ($Q\gg 1$)
\begin{equation}\label{eq:n2d_0}
  n^{(2)}_\text{G}(0)= n_\text{cav}^2 \gamma^2 \ln^2(2Q)
\end{equation}
In the same limit of large quality factors, we can send the lower limit of
integration in Eq.~\eqref{eq:n_d0} to $-\infty$ as the integral is cut-off by
the impedance. In this approximation,  $n_\text{G}^{(2)}$ is independent of
the applied voltage bias (which has been effectively sent to infinity). For
large quality factors, the photon-photon correlations assumes the form
\begin{equation}\label{eq:n2_appr}
  n^{(2)}_\text{G}(\tau) = n_\text{cav}^2 \gamma^2
  \begin{cases}
    16 (\gamma \tau)^{-4}, & \gamma \tau \gg 1, \\
    \ln^2(\gamma \tau), & \gamma \tau \ll 1.
  \end{cases}
\end{equation}
This shows that the correlation time is given by $\gamma$. As we have
neglected the finite value of the voltage, the results in \eqref{eq:n2_appr}
are only valid in the regime $Q^{-1} \lesssim \gamma \tau \lesssim Q$.

Outside this regime, we have to include the finite value of $e V _0= \hbar
\omega_0$.  For small times ($\gamma \tau < Q^{-1}$), the weak divergence for
$\tau\to0$ has to be cut-off at $ \omega_0\tau \simeq 1$. For long times
($\gamma \tau > Q$), the photon-photon correlator shows an oscillatory
component, cf.\
 Fig.~\ref{fig:n2}. The oscillatory part is approximately given by
\begin{equation}\label{eq:n2_osc}
  n^{(2)}_\text{G}(\tau) = \frac{ 8 n_\text{cav}^2 [1-
   \sin( e V_0
 \tau /\hbar) ]}{\omega_0 \tau^3},
\end{equation}
valid for $\omega_0 \tau \gtrsim 1$. The reason for this oscillation lies in
the fermionic statistics of the electrons producing the radiation. At low
temperatures, due to the Pauli principle, the electrons contributing to the
charge transport are separated by a time-interval $\hbar/eV_0$ that is a
remnant of the exchange-hole found in a Fermi sea.\cite{martin:92,hassler:08}
The fact that electrons with the same spin are separated from each other
leads to a separation of the photons which are produced by the electrons and
correspondingly in a dip of the photon-photon correlator at the relevant
timescale. Note that for long-times where the oscillatory behavior becomes
visible, the photon-photon correlator is a factor $Q^{-2}$ smaller than at
$\tau=0$ such that this oscillation although of theoretical interest will most
likely remain experimentally unobservable.

We continue by discussing  the competing negative contribution
$n^{(2)}_\text{nG}$. For large quality factors $Q\gg 1$,  the impedance
$Z^{(2)}$ concentrates the integral \eqref{eq:n_e0} at values $\omega_{1/2}
\simeq \omega_0$ and $\nu \simeq 0$. The value for $\tau =0$ is approximately
given by
\begin{equation}
  n^{(2)}_\text{nG}(0) = - \frac{4 \pi^2 n_\text{cav}^2\gamma^2 \ln(Q)}{3}.
\end{equation}
For finite times $\tau$, we again proceed by sending the lower limit of
integration to $-\infty$ which corresponds to integrating over the region
$\nu/2 < \omega_1 < \omega_2 < \omega_0 -\nu/2$. This approximation renders
the result (valid for $Q \gtrsim\gamma \tau \gtrsim Q^{-1}$)
\begin{equation}\label{eq:n2_eappr}
  n^{(2)}_\text{nG}(\tau) = - \frac{2\pi^2 n_\text{cav}^2 \gamma^2}{3}
  \begin{cases}
    3 (\gamma\tau)^{-2} , & \gamma \tau \gg 1, \\
    2 |\ln(\gamma \tau)|, & \gamma \tau \ll 1.
  \end{cases}
 \end{equation}
independent of the value of the bias voltage.

\section{Critical temperature from Fano factor}\label{sec:tc}

\begin{figure}[t]
  \centering
  \includegraphics{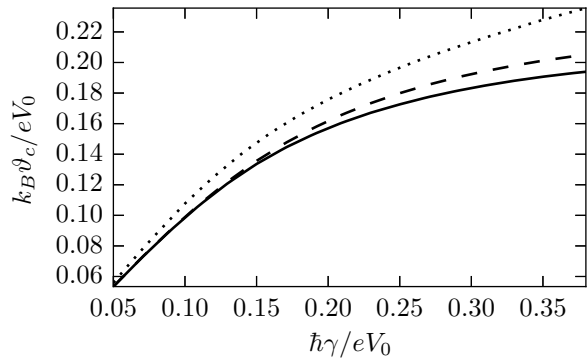}
  \caption{%
    Critical temperatures $\vartheta_c$ and $\vartheta_c^*$   as a function of
    the inverse quality factor $Q= \omega_0/\gamma= eV_0/\hbar \gamma$. The
    dashed line is the evaluation of $\vartheta_c$ with all terms contributing
    to the diagrams in Fig.~\ref{fig:diag}(b) taken into account. This result
    should be compared to the solid line that involves the approximate
    expressions \eqref{eq:nej}.  The results differ by at most a few
    percent showing  that \eqref{eq:nej} is a reasonable approximation for
    the relevant quality factors $Q \gtrsim 5$.  Whereas $\vartheta_c$
    indicates for which temperatures nonclassical signatures of the radiation
    can be detected in the time-averaged quantity (Fano factor), the dotted
    line shows the result for $\vartheta_c^*$, the critical temperature for
    which the second-order coherence $g^{(2)}(\tau)$ is below 1. The latter is
    evaluated with the approximate expressions of Eq.~\eqref{eq:nej}.
  }\label{fig:tc}
\end{figure}

On the one hand, we have seen that for a single channel wire and for large
quality factors, the Fano factor is below 1 at zero temperature, cf.\
Eq.~\eqref{eq:fano}. On the other hand, for large temperatures  $k_B \vartheta
\gg e V_0$, the Fano factor approaches $F = 1 + n_\text{cav} > 1$,
characteristic for a thermal source.  Thus, there is some critical temperature
$\vartheta_c$ at which the Fano factor is 1. In the time-averaged
quantities, the nonclassical signatures of the radiation source thus is only
visible for $\vartheta < \vartheta_c$. Figure~\ref{fig:tc} shows the result
for $\vartheta_c$ as a function of the inverse quality factor. For $Q \geq
10$, the critical temperature is well-approximated by $\vartheta_c \approx
\hbar \gamma/ k_B$. Note that the fact that the critical temperature scales
with the photon-loss rate $\gamma$ has already been noted in
Ref.~\onlinecite{fulga:10}. For $Q=5$, we have approximately $\vartheta_c
\approx 0.16\, e V_0/k_B$.

\section{Critical temperature from second-order coherence}\label{sec:tcs}

\begin{figure}[t]
  \centering
  \includegraphics{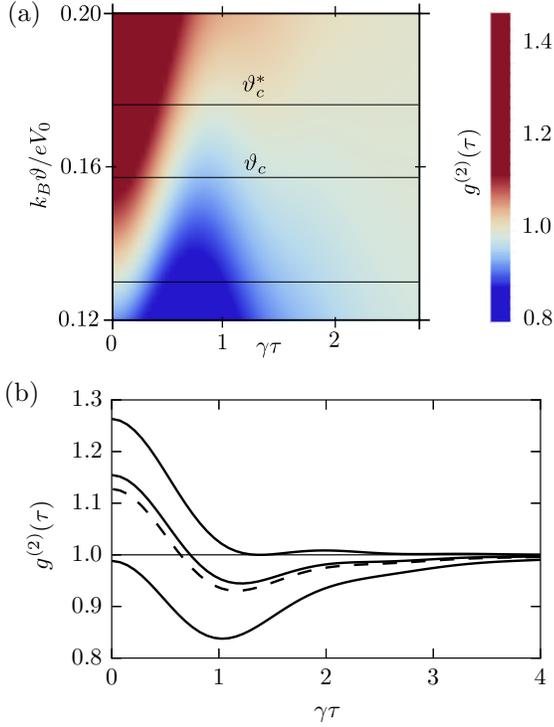}
  \caption{%
    (a) (color online) The second-order coherence $g^{(2)}(\tau)$ as a
    function of the autocorrelation time $\tau$ and the temperature
    $\vartheta$ of the electronic system.  The data have been evaluated for a
    quality factor $Q=5$ using the approximate expressions \eqref{eq:nej}. (b)
    Line cuts (solid lines) for temperatures $\vartheta^*_c$, $\vartheta_c$,
    and $\vartheta = 0.13\, e V_0/k_B$ from top to bottom. The dashed line is
    a calculation for $\vartheta =\vartheta_c$ using the exact expression for
    the diagrams in Fig.~\ref{fig:diag}(b).   It can be seen that the
    approximation \eqref{eq:nej} captures all essential features and even
    underestimates the nonclassical correlations.
  }\label{fig:g2}
\end{figure}

Knowing the critical temperature, we have confirmed numerically  that the
following approximations holds at low temperatures: $n_\text{G}^{(2)}$ remains
largely unchanged with temperature whereas $n_\text{nG}^{(2)}$ scales with a
factor $(1- 0.6 \vartheta/\vartheta_c)$.  As a result, we have the approximate
expression
\begin{equation}\label{eq:finite_temp}
  n^{(2)}(\tau) = n_\text{G}^{(2)} (\tau; \vartheta\!=\!0) + \biggl(1- \frac{0.6
  \vartheta}{\vartheta_c}\biggr)
n_\text{nG}^{(2)} (\tau; \vartheta\!=\!0)
\end{equation}
for the photon-photon correlator at finite temperatures. In Fig.~\ref{fig:n2},
we can see that $ n_\text{G}^{(2)} (\tau; \vartheta\!=\!0)$ falls off fast for
small times when compared to $n_\text{nG}^{(2)} (\tau; \vartheta\!=\!0)$. We
can understand this from the analytical expressions in Sec.~\ref{sec:zero} by
comparing the behavior of $\ln^2|\gamma\tau|$ for $n^{(2)}_\text{G}$ to
$\ln|\gamma \tau|$ for $n^{(2)}_\text{nG}$. Starting from a finite value at
$\tau=0$, $n^{(2)}_\text{G}$ has a point of inflection just  before going over
to the  $\ln^2|\gamma\tau|$ behavior. Numerically, we obtain a value $\tau^*
\approx 5\omega_0^{-1}$ for the position of this point. Due to the fast decay of
$n^{(2)}_\text{G}$ and the fact that the contribution of $n^{(2)}_\text{nG}$
is reduced at finite temperatures, the second-order coherence $g^{(2)}(\tau)$
becomes nonmonotonous at finite temperature with a well-pronounced minimum
slightly above  $\tau^*$, see Fig.~\ref{fig:g2}. Due to the oscillatory
behavior, we expect to obtain $g^{(2)}(\tau \approx \tau^*) <1$ at
temperatures above $\vartheta_c$.  We denote with $\vartheta_c^*$ the critical
temperature below which $g^{(2)}(\tau^*) <1$. The critical temperature  is
shown as the dotted line in Fig.~\ref{fig:tc}. Indeed, for relevant quality
factors $Q<10$, $\vartheta^*_c$ is about 10\% larger than $\vartheta_c$
rendering the requirements to see nonclassical correlation less stringent.

\begin{figure}[t]
  \centering
  \includegraphics{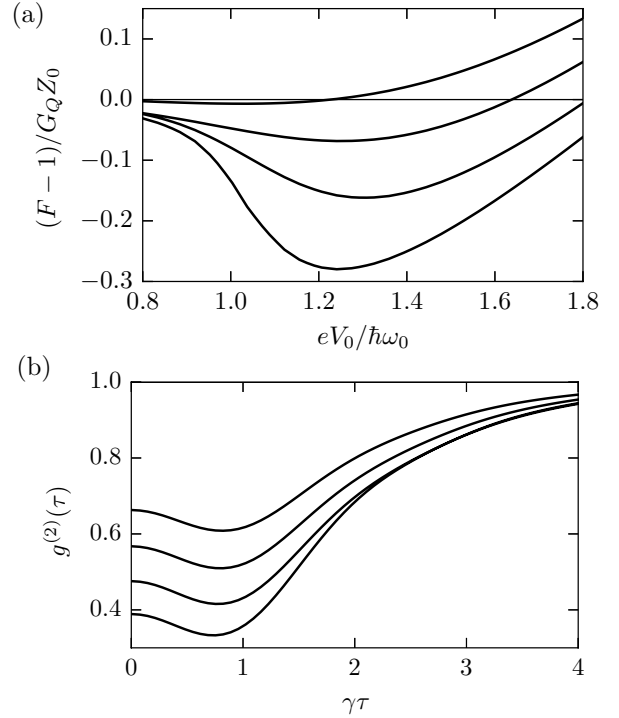}
  \caption{%
    (a) The plot shows the Fano factor $F$ as a function of $V_0$ for
    temperatures $\vartheta= (0,0.05,0.1,0.15) \hbar \omega_0/k_B$ from bottom
    to top. It can be seen that at low temperatures the minimal value of $F$
    is achieved slightly above the resonant condition with $eV_0 \approx
    (1.2$--$1.3) \hbar \omega_0$. (b) Second-order coherence $g^{(2)}(\tau)$
    for the temperature $\vartheta = 0.05 \hbar \omega_0/k_B$ as a function of
    $\tau$.  The different lines correspond to different bias voltages with
    $V_0 = (1.0,1.1,1.2,1.3) \hbar \omega_0/e$ from bottom to top. The
    second-order coherence offers additional time-resolved information when
    compared to the Fano factor. For finite temperatures, $g^{(2)}$ is a
    nonmonotonous function with a minimum at time $\tau^* \approx
    5\omega_0^{-1}$.  All the results have been obtained using the  realistic
    value $Q=\omega_0/\gamma=5$  for the quality factor.
  }\label{fig:volt_tau}
\end{figure}

\section{Experimental Parameters}\label{sec:experimental}

In this section, we would like to present realistic experimental parameters to
observe sub-Poissonian statistics ($F<1$) of the radiation emitted by the
quantum point contact. As can be seen in Fig.~\ref{fig:tc}, the critical
temperature increases with decreasing quality factor. It flattens out for
$\hbar\gamma/eV_0=0.2$.  However, the results have assumed the impedance of
the transmission line to be constant over the range set by $\gamma$.  Thus, a
quality factor of $Q=5$ seems to offer an appropriate balance between having a
large critical temperature while still onnly requiring a moderate bandwidth
for the detector. Since the results of this paper rely on having a single
electronic channel without spin-degeneracy, a point contact in a quantum Hall
edge channel in a sufficiently large magnetic field is essential. The
constricting has to be tuned to a transparency of $T=\tfrac12$. It is
reasonable to expect that the conductance remains linear for voltages up to
$V_0 = 100\,\mu$V. In order to optimize the Fano factor, we propose
$eV_0/\hbar \omega_0 =1.3$ such that $\omega_0 \approx 2\pi\times 18\,$GHz.
In order to achieve $k_B \vartheta/\hbar \omega_0 = 0.05$, an electronic
temperature of the order of $\vartheta \approx 50\,$mK has to be obtained.
From Fig.~\ref{fig:volt_tau}, we can then see that in this case the Fano
factor reads $F= 1 -0.15 G_Q Z_0$. The possibility to observe the
sub-Poissonian nature of the radiation in the end relies on the impedance
matching of the transmission line to the quantum point contact that is
captured in the expression $G_Q Z_0$. For example, if the sensitivity allows
to distinguish $F=1$ from $F=0.95$, an impedance matching with $G_Q Z_0= 0.33$
($Z_0 \approx 8\,$k$\Omega$) is required.

\section{Conclusions}

In conclusion, we have calculated the second-order coherence of microwave
radiation produced by a quantum point contact with a single transport channel.
We have obtained the approximate analytical expression Eq.~\eqref{eq:nej} that
provides accurate results for the relevant temperatures where the radiation is
nonclassical. We have shown that at low temperatures and at transparency
$T=\tfrac12$, two photon processes are suppressed due to the cancellation of
two competing terms. As a result, the stringent requirements on the quality
factor of the \emph{LC}-resonator can be relaxed which helps to increase the
measurement signal.  We have given explicit analytical results for the
photon-photon correlators at zero temperature and an approximate expression
valid at finite but small temperatures.  We have shown that the second-order
coherence $g^{(2)}(\tau)$ shows a nonmonotonous behavior with am minimum
close to $\tau^*\approx 5 \omega_0^{-1}$.  Taking the  minimum of $g^{(2)}$ as
the criterion for nonclassical radiation, the critical temperature below which
nonclassical features can be observed is increased by 10\% compared to
time-averaged quantities.  We have presented a set of realistic though
challenging experimental parameters that allow for the detection of the
sub-Poissonian radiation emitted by the quantum point contact.

\vspace*{1cm}

\acknowledgments

The authors thank Fabien Portier for fruitful discussions and for providing
the motivation for this research. They acknowledge financial support via the
Alexander von Humboldt-Stiftung.

\end{document}